\documentstyle[aps,epsfig,float]{revtex}

\begin{document}
\title{Theory of pairing in the Cu-O Plane: Three-Band Hubbard Model
and Beyond }
\author{Michele Cini, Gianluca Stefanucci and Adalberto Balzarotti}
\address{INFM, Dipartimento di Fisica,Universit\`{a} di Roma Tor Vergata, Via\\
della Ricerca Scientifica 1- 00133 Roma, Italy}
\date{\today}
\maketitle

\begin{abstract}
We calculate the effective interaction $W_{eff}$ between two holes added to the
ground state of the repulsive three-band Hubbard model.  To make contact with Cooper theory and with
earlier Hubbard model cluster studies, we first use a perturbative
canonical transformation, to generate a two-body Hamiltonian. Then, we extend the results to all
orders. The approach is exact in principle, and we obtain a close analytic expression including
explicitly the effects of all virtual transitions to 4-body intermediate states.
Our scheme naturally lends itself to embody off-site, inter-planar, phonon-mediated and other
interactions which are not considered in the Hubbard model but may 
well be important.
The result depends qualitatively on the symmetry of the two-hole state:
$^{1}B_{2}$ and $ ^{1}A_{2}$ pairs are special, because the bare holes do not
interact by the on-site repulsion (W=0 pairs). The effective interaction in these
channels is attractive and leads to a Cooper-like instability of the Fermi
liquid; however $W_{eff}$ is repulsive for Triplet pairs. Bound two-hole states
of the same nature were reported earlier in small cluster calculations by exact
diagonalisation methods; only symmetric clusters are good models of 
the plane.  Once $W_{eff}$  is known, the pair eigenfunction is 
determined by an integral equation. We
present numerical estimates of the binding energy 
$|\Delta|$ of the pairs, which is in the physically interesting range of tens of meV 
if unscreened on-site repulsion parameters are used.
\end{abstract}


\noindent

\narrowtext
\twocolumn
\section{Introduction}

Cooper pairs in high-$T_{C}$ superconductors\cite{kn:bm} are more tighly bound
and  spatially localized than in ordinary BCS ones, and this happens in the
presence of strong correlations. If phonons are mainly responsible for the
mechanism, an enhanced electron-phonon vertex must be involved\cite{kn:pietro} and
a polaron or bipolaron behavior\cite{kn:iad} is also a serious possibility. 
However the wave function of the hole system must be such that the repulsion
barrier is minimized and can be overcome by phonon exchange. Several Authors
have pushed the argument further, proposing that correlation effects alone could
be so strong to turn the repulsion into attraction by some sort of
overscreening and lead to pairing. The proposed electronic mechanisms have often
been based on Hubbard\cite{kn:bob,kn:pw} or t-J models\cite{kn:sus}, just to
cite a few of the less exotic proposals. The phonon-based and electronic
mechanisms have usually been regarded as alternative, but there is a clear need
for a theoretical framework where they can coexist and can be compared on equal
footing; in our opinion, all the serious mechanisms must convey part of the
physics.

Our first aim here is to propose such a scheme, starting with the three-band Hubbard
model. The Hamiltonian is 
\begin{equation}
H=H_{0}+W  \label{tbhm}
\end{equation}
and the independent hole term reads, in the site representation

\begin{equation}
H_{0}={\sum_{Cu}}\varepsilon _{d}n_{d}+{\sum_{O}}\varepsilon _{p}n_{p}+{\
t\sum_{n.n.}}\left[ c_{p}^{\dagger}c_{d}+h.c.\right]  \label{h0}
\end{equation}

where $n.n.$ stands for nearest neighbors. The on-site repulsion term
will be denoted by 
\begin{equation}
W={\sum_{i}}U_{i}n_{i+}n_{i-};  \label{w}
\end{equation}
where $U_{i}=U_{d}$ for a Cu site, $U_{i}=U_{p}$ for an O.

Several years ago, pioneers in cluster 
calculations\cite{kn:hirsch,kn:bal,kn:jones}
explored the possibility that pairing could result from repulsion. They proposed
the following definition\cite{kn:hirsch} of the energy of the pair:

\begin{equation}
\Delta=E(N+2)+E(N)-2E(N+1),  \label{delta}
\end{equation}
where $E(N)$ is the ground state energy of the cluster with $N$ holes, as
obtained by exact diagonalisation. A negative $\Delta$ means that the
ionisation potential decreases with increasing the number of holes, and this
may be taken as an indication of pairing. Several clusters were studied for $N=2$.
It turned out that $\Delta<0$ is possible only if the off-site $U$ between Cu and O is
taken to be unphysically large (about 5 eV).  Moreover, the physical
interpretation was not obviuos, and there was no clear-cut reason for 
excluding hole bags. Also, interestingly, S. Mazumdar et al.\cite{kn:tink}
argued that $\Delta<0$ could be an artifact due to the neglect of the degrees
of freedom of the nuclei: when they are left free to move, the state 
with $N+1$
holes could gain enough energy from Jahn-Teller distortion that
$\Delta$ could turn out to be positive, after all.
Later, we pointed out\cite{kn:cb1} that a couple of key ingredients were missing,
namely, symmetry and W=0 pairs. W=0 pairs are two-hole eigenstates of the kinetic
energy $H_{0}$ that are also eigenstates of the on-site repulsion term $W$ with
eigenvalue 0. There is no on-site repulsion barrier to overcome in W=0 pairs.

 The symmetry of the cluster is essential, because 
 only clusters with the same point symmetry group $C_{4v}$ as the plane
 and centered on a Cu ion (fully symmetric clusters)  allow
such solutions. Still, also the planar lattice structure is
 essential, because no W=0 pairs occur in 3D or in a continuous model, where the interaction term
 (\ref{w}) becomes a contact term.  We started\cite{kn:cb1} by the above
definition of $\Delta$, and studied the fully symmetric clusters
with up to 21 atoms by exact diagonalisation; $\Delta$ was found to be negative
when (and only when) the least bound holes formed a W=0 pair. The Hamiltonian had
been parametrised by electron spectroscopy studies\cite{kn:saw} and 
 ab-initio calculations\cite{kn:sch},
and we used literature values. We also considered first-neighbor off-site
interactions\cite{kn:cb1}. The O-O hopping can change the order of 
single-hole levels, thereby changing the occupation number which is 
necessary to get a partially occupied degenerate state; however, when 
the conditions are satisfied, the $\Delta<0$ behavior results without 
important modifications. The first neighbor off-site Cu-O interactions 
were found to enhance the effect; the first neighbor off-site 
O-O interactions unpair it, but their effect is small if the recommended 
values of the parameters are used. In the 
present paper, such terms are dropped for the sake of simplicity, 
since they are not essential for our present scope. Recent 
calculations by Sch\"{u}ttler et al\cite{kn:schu}, based on a 
combination of diagrammatic and Quantum Monte Carlo methods on an 
Extended Hubbard Model, support the present view that the screened 
interaction is attractive and the attraction is robust against the 
long-range part of the Coulomb repulsion.

In the fully symmetric Cu-O clusters a genuine pairing takes place, due to an effective interaction
which is attractive for singlets and repulsive for triplets.  Let us summarize 
why  this conclusion is free from the ambiguity pointed out in Ref.\cite{kn:tink}.

First, we
computed $\Delta$ for our clusters by second-order ground state energy diagrams
(modified for degenerate ground states, when appropriate)\cite{kn:cb5}. It is convenient to
denote the hole orbitals by their symmetry labels, with $b$ for $b_{1}$ and $a$,
($a^{\prime}$) for the occupied (empty) orbitals of $a_{1}$ symmetry: we obtained
\begin{equation}
\Delta ^{(2)}=-2\left[ 
\mathop{\displaystyle \sum }%
\limits_{b}^{{}}\frac{W(a,b,x,x)^{2}}{(\varepsilon _{b}-\varepsilon _{a})}-%
\mathop{\displaystyle \sum }%
\limits_{a^{\prime}}^{{}}\frac{W(a,a^{\prime },x,x)^{2}}{(\varepsilon _{a^{\prime
}}-\varepsilon _{a})}\right] ,  \label{clust}
\end{equation}
where $\varepsilon _{a}$ is the one-hole energy of the $a$ orbital and so
on; the sums run over all empty states of the appropriate symmetries and
involve the matrix elements of the on-site interaction $W.$ No contributions
arise to second order from the empty states of $e$ symmetry since the
relevant matrix elements vanish.

Second, we computed to second-order the two-hole amplitude for holes of opposite
spins in the degenerate $(x,y)$ orbitals. The only contributions are the
second-order diagrams of Figure 1.
\begin{figure}[H]
\begin{center}
	\epsfig{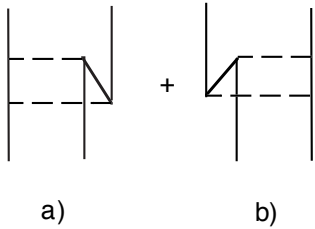}\caption{\footnotesize{
	The second order diagrams for the Two-Hole Amplitude.}}
\end{center} 
\end{figure}

The system makes virtual transitions to 4-body
(3 holes and 1 electron) states. We demonstrated that this produced an effective
interaction, which pushes down the singlet and up the triplet by $| \Delta
^{\left( 2\right) }| $.  In this way, $\Delta$ can be redefined without
any reference to the ground state of clusters with a different number of
holes, and we are free from the objections based on the Jahn-Teller distortion
of odd-$N$ clusters. Good agreement between the second-order calculation and the
numerical exact diagonalisation results supported the interpretation.  Thus, the
cluster calculations\cite {kn:cb2,kn:cb3,kn:cb4} showed that W=0 pairs are the
$^{\prime
\prime }bare^{\prime \prime }$ quasiparticles that, when
$^{\prime\prime }dressed^{\prime \prime }$, become a bound state. 
That approach is
inherently limited by the small size of solvable clusters, but allows a very
explicit display of paired hole properties, that even show superconducting
flux-quantization\cite{kn:cb5,kn:cbs}.  Our second task here is to extend the
theory to the full plane, following the hint provided by the diagrams. In Sect.II
we review the one-body properties of the model, in order to estabilish some
notation and several useful formulas. In Sect.III we show how W=0 pairs in the
full plane can be obtained for any filling. The effective interaction between
holes is calculated by second-order perturbation theory in Sect. IV, with the aim
of showing the relation of the present formalism to Cooper theory from one side
and to the cluster results from the other. With this preparation, we shall be
ready to present in Sect.V the general canonical transformation leading to the
effective Hamiltonian for pairs\cite{kn:csbcond}. This is free from the limitations of
perturbation theory and differs from the results of Sect. IV in an important way,
although it is seen to reduce to them in the appropriate limit. It will turn out
that the effective interactions are additive, and the theory is able to
accommodate the effects of phonons, off-site interactions\cite{kn:cb1},
interplanar coupling\cite{kn:san} and whatever physics one wants to include. The
final result is an integral equation for pairs in the full plane, which is
suitable for a numerical study.  Sect. VI is devoted to the method we used to
solve the integral equation in finite supercells and than taking the asymptotic
limit. Two kinds of bound states of different symmetries result, and Sect. VII
presents the dependence of $\Delta$ on  the filling and other parameters.
Finally, Sect. VIII summarizes our conclusions.

\section{Band Structure}

The eigenstates of

\begin{equation}
H_{0}\psi ^{\left( \nu \right) }\left( k,r\right) =\varepsilon ^{\left( \nu
\right) }\left( k\right) \psi ^{\left( \nu \right) }\left( k,r\right)
\end{equation}
where $\nu $ is a band index, are written according to Bloch's theorem

\begin{equation}
\psi ^{\left( \nu \right) }\left( k,r\right) =e^{ik r}\phi ^{\left( \nu
\right) }\left( k,r\right) ,
\end{equation}
where $\phi $ is periodic. If $r=0$, $r=a$ and $r=b$ are the positions of Cu
and of the two Oxygens, respectively, the non-bonding band is characterized
by $\phi ^{\left( nb\right) }\left( k,0\right) =0;$ one then finds

\begin{equation}
\phi ^{\left( nb\right) }\left( k,a\right) =\frac{\cos \left[ \frac{k_{x}d}{2%
}\right] }{\sqrt{N_{C}\left( \cos ^{2}\left[ \frac{k_{x}d}{2}\right] +\cos
^{2}\left[ \frac{k_{y}d}{2}\right] \right) }},
\end{equation}

\begin{equation}
\phi ^{\left( nb\right) }\left( k,b\right) =\frac{\cos \left[ \frac{k_{y}d}{2%
}\right] }{\sqrt{N_{C}\left( \cos ^{2}\left[ \frac{k_{x}d}{2}\right] +\cos
^{2}\left[ \frac{k_{y}d}{2}\right] \right) }},
\end{equation}
where $N_{C}$ is the number of cells in the crystal and $d$ the lattice
parameter. For the other bands, one obtains the eigenvalue equation

\begin{equation}
\varepsilon ^{\left( \pm \right) }\left( k\right) =\frac{
\varepsilon _{p}\pm r}{2}\label{scorc}
\end{equation}
(+ for the antibonding and - for the bonding band), with

\begin{equation}
r =
\sqrt{\varepsilon _{p}^{2}+16t^{2}\left[ \cos ^{2}\left[ 
\frac{k_{x}d}{2}\right] +\cos ^{2}\left[ \frac{k_{y}d}{2}\right] \right] }
\label{rad}
\end{equation}

 and the Cu amplitudes 
\begin{equation}
\phi ^{\left( \pm \right) }\left( k,0\right) =\sqrt{\frac{\varepsilon
^{\left( \pm \right) }-\varepsilon _{p}}{N_{C}\left( 2\varepsilon ^{\left(
\pm \right) }-\varepsilon _{p}\right) }},
\end{equation}

while the O amplitudes are obtained from

\begin{eqnarray}
\left( \varepsilon _{p}-\varepsilon ^{\left( \nu \right) }\left( k\right)
\right) \phi ^{\left( \nu \right) }\left( k,a\right)\nonumber\\
 +2t\cos \left[ \frac{%
k_{y}d}{2}\right] \phi ^{\left( \nu \right) }\left( k,0\right) =0,
\end{eqnarray}
and
\begin{eqnarray}
\left( \varepsilon _{p}-\varepsilon ^{\left( \nu \right) }\left( k\right)
\right) \phi ^{\left( \nu \right) }\left( k,b\right) \nonumber\\
+2t\cos
\left[ \frac{ k_{x}d}{2}\right] \phi ^{\left( \nu \right) }\left(
k,0\right) =0.       \label{}
\end{eqnarray}

Setting $x=k_{x}d$, $y=k_{y}d$, we get the contours of constant energy
$\varepsilon$ in the Brillouin Zone (BZ):

\begin{equation}
\cos \left( x\right) +\cos \left( y\right) =q\left( \varepsilon \right) ,
\end{equation}

\begin{equation}
q\left( \varepsilon \right) =\frac{\varepsilon -\varepsilon _{p}}{2t^{2}}-2.
\end{equation}

In the bonding band, $q=2$ at the bottom and $q=-2$ at the top; at 
$q=0$ the
contour is a square with vertices at $\frac{2\pi }{d}\left( \pm 1,0\right) $
and $\frac{2\pi }{d}\left( 0,\pm 1\right) $, where the Fermi velocity 
vanishes. Let $D$ be the domain defined by $
\varepsilon < E_{F}$ in the BZ.

Below half filling , 
\begin{equation}
\int\int_{D} dxdy=\int_{{a(0)}}^{-{a(0)}}dx\int_{{a(x)}}^{-%
{a(x)}}dy,
\end{equation}
where ${a(x)=\arccos(q(E_{F})-\cos(x))}$; at half filling 
\begin{equation}
\int\int_{D} dxdy=\int_{-\pi}^{\pi}dx\int_{-\pi+\left| x\right|}^{\pi-\left|
x\right|}dy,
\end{equation}
and above half filling, putting
 $b(x)=\arccos(q(E_{F})+\cos(x))$ , 
we find

\begin{equation}
\int\int_{D} dxdy=I_{1}+I_{2}+I_{3} ,
\label{intot}
\end{equation}
where

\begin{equation}
I_{1}=\int_{-{b(0)}}^{{b(0)}}dx\int_{-\pi}^{\pi}dy ,
\label{i1}
\end{equation}

\begin{equation}
I_{2}=\int_{-\pi}^{-b(0)}dx\int_{-a(x)}^{a(x)}dy ,
\label{i2}
\end{equation}
and
\begin{equation}
I_{3}=\int_{b(0)}^{\pi}dx\int_{-a(x)}^{a(x)}dy ,
\label{i3}
\end{equation}

The number of holes per spin per unit cell below $E_{F}$ in the bonding band
above half filling is

\begin{equation}
N(E_{F})=\left( \frac{d}{2\pi }\right) ^{2}\int\int d^{2}k\theta\left( 
E_{F}-\varepsilon\left( k \right)\right)
\end{equation}
differentiating, one finds the density of states

\begin{equation}
\rho \left( \varepsilon \right) =\frac{\varepsilon _{p}-2\varepsilon }{2\pi
^{2}t^{2}}\int_{\arccos \left[ q+1\right] }^{\pi }\frac{dx}{\sqrt{1-\left(
q-\cos \left( x\right) \right) ^{2}}},
\end{equation}
where $q=q\left( \varepsilon \right) .$ Taking care of the singularity of
the integrand at the lower limit, this expression is convenient for the
numerical evaluation, and shows a logarithmic singularity at half 
filling (Van Hove singularity).

\section{W=0 Pairs}
We change representation with

\begin{equation}
c_{\sigma}^{\dagger}\left(r_{i}\right)=\sum_{k,\nu}c_{k,\nu,\sigma}^{\dagger}
\psi_{\sigma}^{\left( \nu\right)*}\left( k,r_{i} \right),  \label{ccroce}
\end{equation}
where $k$ runs over the BZ and $\nu$ over bands. Thus,

\begin{eqnarray}
W=\sum_{k_{1}..k_{4}}\sum_{\nu_{1}..\nu_{4}}
c_{k_{1},\nu_{1},+}^{\dagger}c_{k_{2},\nu_{2},+}
c_{k_{3},\nu_{3},-}^{\dagger}c_{k_{4},\nu_{4},-}\nonumber\\
\times
U^{\left(\nu_{1}..\nu_{4}\right)}\left( k_{1},k_{3},k_{2},k_{4} \right), 
\label{wk}
\end{eqnarray}
Using a shorthand notation, where the band indices are understood,

\begin{eqnarray}
U\left(m,n,p,q\right)=\sum_{i}
e^{-i\left(m+n-p-q\right)r_{i}}\nonumber\\
\times
\phi^{*}\left( m,i \right)\phi^{*}\left( n,i \right)\phi\left( p,i
\right)\phi\left( q,i \right), 
\label{uk}
\end{eqnarray}
is the lattice Fourier transform of a periodic function; the momenta are
limited to the BZ, so, introducing reciprocal lattice vectors $G$ we may write

\begin{eqnarray}
U\left(m,n,p,q\right)=N_{C}\sum_{G}
\delta\left(m+n-p-q-G\right)\nonumber\\
\times U^{C}\left(m,n,p,q;G\right), 
\label{ug}
\end{eqnarray}
where $U^{C}$ involves a sum over a single cell:

\begin{eqnarray}
U^{C}\left(m,n,p,q;G\right)=\sum_{i}^{Cell}U_{i}
e^{-iG r_{i}} \nonumber\\
\times
\phi\left( m,i \right)\phi\left( n,i
\right)\phi\left( p,i
\right)\phi\left( q,i \right), 
\label{uc}
\end{eqnarray}
where stars have been dropped since $\phi^{\prime}s$ are real.The umklapp
processes involving the basis vectors of the reciprocal lattice   produce a
- sign on one of the Oxygens. The matrix elements (\ref{ug}) are 
proportional to $N_{C}^{-1}$, like any {\em bona fide} two-body 
interaction.

Still omitting the band indices, we shall mean 
\begin{equation}
d[k]=\left\| k_{+},-k_{-}\right\| =c_{k,+}^{\dagger}c_{-k,-}^{\dagger}|vac> 
\label{3}
\end{equation}
to be a two-hole determinantal state derived from the $k$ eigenfunctions.
Since $\phi\left(-k,r\right)=\phi\left(k,r\right)$, which is required by
time-reversal symmetry, the combination $d[k]+d[-k]$ is singlet and
$d[k]-d[-k]$ is triplet. 

The point symmetry Group of the Cu-O plane is $C_{4v}$, and its character
Table is shown in Table I.

We introduce the determinants
$Rd[k]=d[Rk]=d[k_{R}],R\in C_{4v}$ , and the projected states 
\begin{equation}
\Phi _{\eta }\left[ k\right] =\frac{1}{\sqrt{8}}{\sum_{R\in C_{4v}}}\chi
^{\left( \eta \right) }\left( R\right) Rd[k]  \label{4}
\end{equation}
where $\chi ^{\left( \eta \right) }(R)$ is the character of the operation $R$
in the Irreducible Representation (Irrep) $\eta $. In the non-degenerate
Irreps, the operations that produce opposite $k_{R}$ have the same character,
and the corresponding projections lead to singlets. Let $R_{i},i=1,..8$
denote the operations of $C_{4v}$ and $k,k^{\prime} $ any two points in the BZ.
Consider any two-body operator $\hat{O}$, which is symmetric
($R_{i}^{\dagger}\hat{O}R_{i}=\hat{O}$), and the matrix with elements $%
O_{i,j}=<d[k]|R_{i}^{\dagger}\hat{O}R_{j}|d[k^{\prime }]>$, where $k$ and $ k^{\prime
}$ may be taken to be in the same or in different bands. This matrix can be
diagonalized by Group Theory. Indeed, for each Irrep
$\eta$, consider the normalized vector with components 
$\frac{1}{\sqrt{8}}\chi ^{\left( \eta \right) }\left( R_{i}\right); $
this is an eigenvector of the operator matrix $R_{i}^{\dagger}\hat{O}R_{j}$, as we
can check by noting that
$
\frac{1}{8}\sum_{i,j}\chi^{\left(\alpha\right)}\left(R_{i}\right)
R_{i}^{-1}\hat{O}R_{j}\chi^{\left(\beta\right)}\left(R_{j}\right)
=\hat{O}P^{2}\left(\alpha\right)\delta\left(\alpha,\beta\right);$
hence, $R_{i}^{\dagger}\hat{O}R_{j}$ is diagonal on the basis of the symmetry
projected states.The square of the projection operator is
$P^{2}\left(\alpha\right)=\sqrt{8}P\left(\alpha\right).$
Now, taking the matrix element between the $k$ and $k^{\prime}$  determinants,
 we get the eigenvalues in terms of determinantal matrix elements:
 
\begin{equation}
O\left( \eta ,k,k^{\prime }\right) ={\sum_{R}}\chi ^{\left( \eta \right)
}\left( R\right) O_{R}\left( k,k^{\prime }\right)  \label{os1}
\end{equation}
where 
\begin{equation}
O_{R}\left( k,k^{\prime }\right) =\langle d[k]|\hat{O} |Rd[k^{\prime
}] \rangle .  \label{os2}
\end{equation}
Thus, omitting the $k$, $k^{\prime }$ arguments, we get 

\begin{eqnarray}
O\left( ^{1}A_{1}\right) =O_{E}+O_{C_{2}}+O_{C_{4}}+O_{C_{4}^{3}} 
\nonumber \\
+O_{\sigma _{x}}+O_{\sigma _{y}}+O_{\sigma _{1}^{\prime }}+O_{\sigma
_{2}^{\prime }}  \label{a1}
\end{eqnarray}

\begin{eqnarray}
O\left( ^{1}A_{2}\right) =O_{E}+O_{C_{2}}+O_{C_{4}}+O_{C_{4}^{3}} 
\nonumber \\
-O_{\sigma _{x}}-O_{\sigma _{y}}-O_{\sigma _{1}^{\prime }}-O_{\sigma
_{2}^{\prime }}  \label{7}
\end{eqnarray}
\begin{eqnarray}
O\left( ^{1}B_{2}\right) =O_{E}+O_{C_{2}}-O_{C_{4}}-O_{C_{4}^{3}} 
\nonumber \\
-O_{\sigma _{x}}-O_{\sigma _{y}}+O_{\sigma _{1}^{\prime }}+O_{\sigma
_{2}^{\prime }}  \label{8}
\end{eqnarray}
and for the $x$ component of the degenerate triplet
\begin{equation}
O\left( ^{3}E_{x}\right) =O_{E}-O_{C_{2}}+O_{\sigma _{x}}-O_{\sigma
_{y}}. \label{tripl}
\end{equation} 
Triplets of the other Irreps are obtained from determinants of the form
$\left\|k_{1},k_{2} \right\|$, but they vanish when $k_{2}=-k_{1}$.
If
$\hat{O}$ is identified with $W$, the determinantal matrix element does not
depend on the sign of the components of
$k^{\prime}$, because the
$\phi^{\prime}s$ depend on $k$ only through a cosine. The only thing that
matters is that $C_{4},C_{4}^{3},\sigma_{1}^{\prime}$ and
$\sigma_{2}^{\prime}$ exchange $k_{x}^{\prime}$ and $k_{y}^{\prime}$, while
the other operations leave them in place. Consequently, 
$W_{E}=W_{C_{2}}=W_{\sigma _{x}}=W_{\sigma _{y}}$ and
$W_{C_{4}}=W_{C_{4}^{3}}=W_{\sigma _{1}^{\prime }}=W_{\sigma _{2}^{\prime
}}$, and 
\begin{equation}
W\left( ^{1}A_{2}\right) =W\left( ^{1}B_{2}\right) =0 .
\end{equation}
These are W=0 pairs, like those studied previously\cite{kn:cb5}, but with an
important change, since in the cluster calculations the symmetries of W=0 pairs
were found\cite{kn:cbs} to be
$^{1}B_{2}$ and $^{1}A_{1}$. The reason for this change is a twofold size effect.
On one hand, $^{1}A_{1}$ pairs have the W=0 property only in the small clusters,
having the topology of a cross, and belonging to the $S_{4}$ Group, but do not
generalize as such to the full plane, when the symmetry is lowered to $C_{4v}$; on
the other hand, the small clusters admit no solutions of
$^{1}A_{2}$ symmetry at all.

 One
necessary condition for pairing in clusters is that the least bound holes form
such a pair, and this dictates conditions on the occupation number. In the full
plane, however, W=0 pairs exist at the Fermi level for any filling. The explicit
W=0 singlet pair states are: 
\begin{eqnarray}
\Phi _{^{1}B_{2}}\left[ k,r_{1},r_{2}\right] =\frac{\chi _{0}}{\sqrt{2}}
\{
\cos\left[ k(r_{1}-r_{2})\right] \phi \left( k,r_{1}\right) \phi \left(
k,r_{2}\right)   \nonumber \\
-\cos\left[ k_{C_{4}}(r_{1}-r_{2})\right] \phi \left( k_{C_{4}},r_{1}\right) \phi
\left( k_{C_{4}},r_{2}\right)\nonumber \\
 -\cos\left[ k_{\sigma}(r_{1}-r_{2})\right] \phi
\left( k_{\sigma },r_{1}\right) \phi \left( k_{\sigma },r_{2}\right)
 \nonumber \\
+\cos\left[ k_{\sigma^{\prime} }(r_{1}-r_{2})\right] \phi
\left( k_{\sigma^{\prime} },r_{1}\right) \phi \left( k_{\sigma^{\prime}
},r_{2}\right)
\}\label{b2e}
\end{eqnarray}
and  
\begin{eqnarray}
\Phi _{^{1}A_{2}}\left[ k,r_{1},r_{2}\right]  =\frac{\chi _{0}}{\sqrt{2}}
\{
\cos\left[ k(r_{1}-r_{2})\right] \phi \left( k,r_{1}\right) \phi \left(
k,r_{2}\right)   \nonumber \\
+\cos\left[ k_{C_{4}}(r_{1}-r_{2})\right] \phi \left( k_{C_{4}},r_{1}\right) \phi
\left( k_{C_{4}},r_{2}\right)\nonumber \\
 -\cos\left[ k_{\sigma}(r_{1}-r_{2})\right] \phi
\left( k_{\sigma },r_{1}\right) \phi \left( k_{\sigma },r_{2}\right)
 \nonumber \\
-\cos\left[ k_{\sigma^{\prime} }(r_{1}-r_{2})\right] \phi
\left( k_{\sigma^{\prime} },r_{1}\right) \phi \left( k_{\sigma^{\prime}
},r_{2}\right)
\},\label{a2e}
\end{eqnarray}
where $\chi_{0}$ is a singlet spin function.
Using the above explicit Bloch states we can verify that both vanish for
$r_{1}=r_{2}$. This is another way to see that they have the W=0 property. All
the distinct W=0 pairs of each symmetry may be labeled from the $k$ points in 1/8
of the empty part of the BZ: for instance, those with $k_{x}>k_{y}>0$. We
shall denote such a set by $e/8$.

\section{Canonical Transformation-Second Order}
In this Section, we shall parallel  as closely as possible the Cooper
theory\cite{kn:kittel} where an effective interaction involving phonons is
introduced via an approximate canonical transformation. The treatment we
give here is based on second-order perturbation theory and is not yet
adequate for the present purposes, but is useful as a prelude to the more
complete theory given in the next Section. We wish to stress the formal
analogy with the BCS theory and make contact with the
$\Delta$ expression derived from cluster studies. Suppose we add two holes to a
background Fermi {\em sphere}. We use Roman indices for the
zeroth-order pair states, which satisfy
\begin{equation}
H_{0}\left| m\right\rangle =E_{m}\left| m\right\rangle
\end{equation}
with $E_{m}=2\varepsilon\left(m\right)$. The basic idea here is that of obtaining
an effective interaction for two holes using the process of the diagrams 
in Fig.1.
Accordingly, the pair states are coupled to the set of 3 hole-1 electron
intermediate states, which satisfies

\begin{equation}
H_{0}\left| \alpha \right\rangle =E_{\alpha }\left| \alpha \right\rangle.
\end{equation}
Thus, the effect of the perturbation is of the form 

\begin{equation}
W\left| m\right\rangle =\sum_{m^{\prime }}\left| m^{\prime }\right\rangle
W_{m^{\prime },m}+\sum_{\alpha }\left| \alpha \right\rangle W_{\alpha 
,m} \label{equa}
\end{equation}
where $W_{m',m}$ stands for the interaction between pairs, while $\alpha $ is
the set of 3 hole-1 electron intermediate states which are coupled to the
pairs by the diagram; 
\begin{equation}
W\left| \alpha \right\rangle =\sum_{m}\left| m\right\rangle W_{m,\alpha },
\end{equation}
The off-diagonal $W_{m',m}$ elements vanish for W=0 pairs; 
on the other hand, in the many-body problem, the diagonal elements $W_{m,m}$ do not 
vanish, because of the tadpole-diagram contributions to the 
self-energy of the holes. However, the diagonal terms do not 
contribute to the effective interaction, but simply renormalize the 
$\varepsilon$ parameters in $H_{0}$. We had already met these tadpole 
diagrams in the calculation of $\Delta^{\left(2\right)}$ for 
clusters\cite{kn:cb5} where they were seen to cancel in the 
calculation of (\ref{delta}). Anyway, we keep $W_{m',m}$ in 
(\ref{equa})  which allows to 
introduce the effects of other interactions which are not included in the Hubbard model.
In the present Section we shall consider $W$ as a small 
 operator and look for an approximate canonical
transformation such that the new Hamiltonian $\tilde{H}$ decouples the 
$\alpha$ states to first order; then, $\tilde{H}$ operates on the space of pairs.
In other terms, we seek a first-order anti-Hermitean operator
$\Lambda$ such that
\begin{equation}
\tilde{H}=e^{-\Lambda}He^{\Lambda}
\end{equation} 
has no linear term connecting the $|\alpha\rangle$ states with the $|m\rangle$
 states. Since
\begin{equation}
\tilde{H}=H+\left[ H,\Lambda \right] +
\frac{1}{2}\left[ \left[
H,\Lambda \right] ,\Lambda \right] 
+\O\left( \Lambda^{3}\right)
\end{equation}
expands to read
\begin{equation}
\tilde{H}=H+\left[ H_{0},\Lambda \right] +\left[ W,\Lambda \right]
+\frac{1}{2}\left[
\left[ H_{0},\Lambda\right] ,\Lambda\right] +...\label{hs}
\end{equation}
 this is accomplished if
\begin{equation}
W_{\alpha ,m}+\left( E_{\alpha }-E_{m}\right) \Lambda_{\alpha,m}=0
\end{equation}
with all others $\Lambda$-matrix elements equal to zero.

  Assuming that the denominators do not vanish (more about that later) 
we obtain

\begin{equation}
\left\langle \alpha \right| \Lambda \left| m\right\rangle =\frac{\left\langle
\alpha \right| W\left| m\right\rangle }{E_{m}-E_{\alpha }}
\end{equation}

In $\tilde{H}$, the holes interact through the effective vertex of Figure 2.

\begin{figure}[H]
\begin{center}
	\epsfig{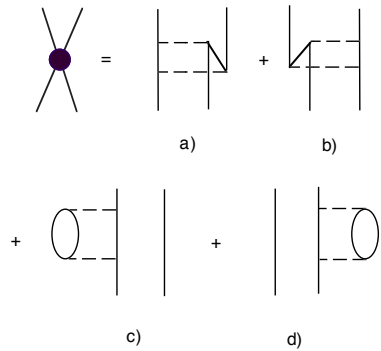}\caption{\footnotesize{
	The effective hole-hole vertex (blob) defined by
the second-oder $\tilde{H}$. The dashed lines represent $W$ interactions. Linked
diagrams belong to $ \tilde{W}_{eff}$, and unlinked ones to $F$.}}
\end{center} 
\end{figure}

The
interaction lines in the right hand side represent $W$; while a) and its specular
counterpart b) correspond to the effective interaction diagrams of Figure 
1, c) and d) are unlinked and represent self-energy processes of each hole.
The same situation occurs in Cooper theory, where the self-energy processes
involve the emission and absorption of a phonon by the same electron.

We may write
\begin{equation}
\tilde{H}=H_{0}+F+\tilde{W}_{eff}\label{hs4}
\end{equation}
where $F$ is diagonal in the pair space, like $H_{0}$, and corresponds to the
unlinked self-energy diagrams, while the effective interaction operator is
$\tilde{W}_{eff}$. Like in Cooper theory, $F$ will be dropped. Formally, we
write

\begin{equation}
F+\tilde{W}_{eff}=\frac{1}{2}\left[W,\Lambda \right] \label{hs5}
\end{equation}
and 

\begin{eqnarray}
2\left\langle m \right| F+\tilde{W}_{eff}\left| m^{\prime}\right\rangle
\nonumber\\
=\sum_{\alpha}W_{m,\alpha}W_{\alpha,m^{\prime}}\left[
\frac{1 }{E_{m}-E_{\alpha }}+\frac{1 }{E_{m^{\prime}}-E_{\alpha }}
\right] \label{fwtilde}
\end{eqnarray}
The $\alpha $ states are 3 hole-1 electron determinants which carry no
quasi-momentum. We write

\begin{equation}
|\alpha >=|\left\| \left( k^{\prime }+q+k_{2}\right) _{+},\bar{k}%
_{2-},-q_{-},-k_{-}^{\prime }\right\| >  \label{alfas}
\end{equation}
where $\bar{k}_{2+}$ is the electron state and pedices refer to the spin
direction; $\alpha$ states with opposite spin indices arise from the specular diagram,
contribute similarly and yield a factor of 2 at the end. The unperturbed
eigenvalues are:
\begin{equation}
E_{\alpha}=\varepsilon\left(k^{\prime
}+q+k_{2}\right)-\varepsilon\left(k_{2}\right)+
\varepsilon\left(q\right)+\varepsilon\left(k^{\prime }\right).
\label{ealfa}
\end{equation}
 Let us first do the
calculation on a determinantal basis. We work out the interaction matrix element:

\begin{eqnarray}
<\left\| \left( k^{\prime }+q+k_{2}\right) _{+},\bar{k}
_{2-},-q_{-},-k_{-}^{\prime }\right\| |W|d[s]>=  \nonumber \\
\delta \left( q-s\right) U\left( q+k^{\prime }+k_{2},-k^{\prime
},s,k_{2}\right)  \nonumber \\
-\delta \left( k^{\prime }-s\right) U\left( q+k^{\prime
}+k_{2},-q,s,k_{2}\right)  \label{me}
\end{eqnarray}
two of the 4 terms in the product in (\ref{fwtilde}) are proportional to
$\delta\left(p-s\right)$ and belong to $F$, while the cross terms yield
identical contributions to the effective interaction. Symbolically, we may
write 
\begin{eqnarray}
\left\langle m\right| \tilde{W}_{eff}\left| m^{\prime }\right\rangle=\frac{
1}{2}\sum_{\alpha }\left[ W_{m,\alpha }W_{\alpha ,m^{\prime }}\right]_{Linked}
\nonumber\\
\left[ \frac{1}{E_{m}-E_{\alpha }}+\frac{1}{E_{m^{\prime
}}-E_{\alpha }}\right]
\label{wtilde}
\end{eqnarray}
Including the specular diagram by a factor of 2, we write the interaction
between determinantal states (with $s$ and $p$ empty).

\begin{eqnarray}
\left\langle d\left[ p\right] \right| \tilde{W}_{eff}\left| d\left[
s\right] \right\rangle =2
\sum_{k}^{occ}\theta \left( \varepsilon \left( s+p+k\right) -E_{F}\right)
\nonumber\\
\times U\left( s+p+k,-p,s,k\right) U\left( p,k,s+p+k,-s\right)
\nonumber\\
\{\frac{1}{\varepsilon \left( s+p+k\right) -\varepsilon \left( k\right)
-\varepsilon \left( s\right) +\varepsilon \left( p\right) } \nonumber\\
+\frac{1}{\varepsilon \left( s+p+k\right) -\varepsilon \left( k\right)
-\varepsilon \left( p\right) +\varepsilon \left( s\right) }\}. \label{wd}
\end{eqnarray}
The interaction between symmetry projected states is obtained using 
(\ref{os1},\ref{os2}) and is given by

\begin{eqnarray}
\left\langle \Phi_{\alpha}\left[ p\right] \right| \tilde{W}_{eff}\left|
\Phi_{\alpha}\left[ s\right] \right\rangle =\nonumber\\
\sum_{R}\chi^{\left(\alpha\right)}\left(R\right)
 \left\langle d\left[ p\right] \left| \tilde{W}_{eff}\right|
Rd\left[ s\right] \right\rangle
\label{wfi}
\end{eqnarray}
The above theory applies not only to the plane but also to clusters, provided that
they are fully symmetric and allow W=0 pairs. The natural basis set is provided
by the orbitals, rather than the Bloch states; in the small clusters with open
boundary conditions that we studied previously, pairing occurs  when a degenerate
state of the Irrep $e(x,y)$ is occupied by two holes. Then, in terms of orbitals,
the only $^{1}B_{2}$ W=0 pair available for pairing is
\begin{equation}
\Phi _{^{1}B_{2}}=\frac{\left\|x_{+}y_{-}\right\|+\left\|y_{+}x_{-}\right\|}{
\sqrt{2}}.
\end{equation}
The diagonal element of $\tilde{W}_{eff}$ on this state gives the second-order
energy shift of the pair; one calculates the $W$ matrix on the orbital basis and 
after some algebra one finds the above $\Delta^{\left(2 \right)}$  expression
(\ref{clust}). In the small clusters, the diagonal element provides
the binding energy, while the localization is granted by the
boundary conditions. In the full plane, there is a continuous infinity of W=0
pairs for each of the allowed symmetries, and the off-diagonal elements are essential
to form a localized wave packet. Therefore the localized state is not 
known {\em a priori}, and  $\Delta$ cannot be obtained as 
an expectation value of $\tilde{W}_{eff}$ , but must be found by solving a two-body problem with 
interaction $\tilde{W}_{eff}$  (See Sect.VI).

In the continuum limit, the integral in (\ref{wfi}) must be understood as a
principal part, because the denominators vanish in a set of vanishing measure
within the integration domain. While in principle this is no trouble,
the numerical evaluation is difficult. However, this computational
difficulty is an artifact of the perturbation approach, as will become clear in
the next Section.

 Above all, the approximate canonical transformation is open to
question because it assumes that $W$ be weak. In small clusters, it works
remarkably well, even for at large couplings, but one could argue that the reason
might be that $\Delta$ is still small compared to the average distance
between the discrete levels. When the levels form a continuum, the
applicability of perturbation theory is not obvious. Therefore, it is important
to put the theory on a clearer and firmer basis, as we do in the next
Section. {\em A posteriori}, it turns out that high order diagrams
mererly renormalize some quantities without destroying the structure
of the theory, because it is a property of W=0 pairs that the interaction is
dynamically small.

\section{Canonical Transformation - All Orders}

Suppose the Cu-O plane is in its ground state with Fermi energy $E_{F}$ and
a couple of extra holes are added. We wish to show that by a
canonical transformation\cite{kn:csbcond} one obtains an effective Hamiltonian which describes
the propagation of a pair of {\em dressed} holes, and includes
{\em all} many-body effects.

The exact many-body ground state with two added holes may be expanded in
terms of excitations over the vacuum (the non-interacting Fermi {\em sphere}) 
by a configuration interaction: 
\begin{equation}
|\Psi _{0}>={\sum_{m}}a_{m}|m>+{\sum_{\alpha }}b_{\alpha }|\alpha >+{\
\sum_{\beta }}c_{\beta }|\beta >+....  \label{psi0}
\end{equation}
here $m$ runs over pair states, $\alpha $ over 4-body states ($2$ holes and $1$
e-h pair), $\beta $ over 6-body ones ($2$ holes and $2$ e-h pairs), and so on.
To set up the Schr\"{o}dinger equation, we consider the effects of the
operators on the terms of $|\Psi _{0}>$. We write:

\begin{equation}
H_{0}|m>=E_{m}|m>,\;H_{0}|\alpha >=E_{\alpha }|\alpha >,...  \label{h0m}
\end{equation}
and since $W$ can create or destroy up to 2 e-h pairs,

\begin{eqnarray}
W|m>={\sum_{m^{\prime }}}W_{m^{\prime },m}|m^{\prime }>+{\sum_{\alpha }}%
|\alpha >W_{\alpha ,m}  \nonumber \\
+{\ \sum_{\beta }}|\beta >W_{\beta ,m}.  \label{wm}
\end{eqnarray}
As explained in the previous Section, $W_{m^{\prime },m}$ does not 
contribute to the effective interaction for W=0 pairs in our model; however we keep it
for generality, since it allows to introduce the effect of the 
exchange of phonons and
other quasiparticles that we are not considering. For clarity let us
first write the equations that include explicitly up to 6-body states; then
we have 
\begin{eqnarray}
W|\alpha >={\sum_{m}}|m>W_{m,\alpha }+{\sum_{\alpha ^{\prime }}}|\alpha
^{\prime }>W_{\alpha ^{\prime },\alpha }  \nonumber \\
+{\sum_{\beta }}|\beta >W_{\beta ,\alpha }  \label{walfa}
\end{eqnarray}
where scattering between 4-body states is allowed by the second term, and

\begin{eqnarray}
W|\beta >={\sum_{m^{\prime }}}\left| m^{\prime }\right\rangle W_{m^{\prime
},\beta }+{\sum_{\alpha }}\left| \alpha \right\rangle W_{\alpha ,\beta } 
\nonumber \\
+{\sum_{\beta^{\prime } }}\left|\beta^{\prime } \right\rangle W_{
\beta^{\prime } ,\beta }  \label{wbeta}
\end{eqnarray}
In principle, the $W_{\beta^{\prime } ,\beta }$ term can be eliminated by
taking linear combinations of the complete set of $\beta $~ states: when
this is done, we get a self-energy correction to $E_{\beta }$ and a
mixing of the vertices, without altering the structure of the
equations. The Schr\"{o}dinger equation yields equations for the
coefficients $a$,$b$ and $c$ 
\begin{eqnarray}
\left( E_{m}-E_{0}\right) a_{m}  \nonumber \\
+{\sum_{m^{\prime }}}a_{m^{\prime }}W_{m,m^{\prime }}+{\sum_{\alpha }}
b_{\alpha }W_{m,\alpha }+{\sum_{\beta }}c_{\beta }W_{m,\beta } =0
\label{eq65}
\end{eqnarray}

\begin{eqnarray}
\left( E_{\alpha }-E_{0}\right) b_{\alpha }  \nonumber \\
+{\sum_{m^{\prime }}}a_{m^{\prime }}W_{\alpha,m^{\prime }}+{\sum_{\alpha
^{\prime }}}b_{\alpha ^{\prime }}W_{\alpha ,\alpha ^{\prime }}+{\sum_{\beta }
}c_{\beta }W_{\alpha ,\beta } =0  \label{eq66}
\end{eqnarray}

\begin{equation}
\left( E_{\beta }-E_{0}\right) c_{\beta }+{\sum_{m^{\prime }}}a_{m^{\prime
}}W_{\beta ,m^{\prime }}+{\sum_{\alpha ^{\prime }}}b_{\alpha ^{\prime
}}W_{\beta ,\alpha ^{\prime }}=0  \label{16}
\end{equation}
where $E_{0}$ is the ground state energy. Then, we exactly decouple the
6-body states by solving the equation for $c_{\beta }$ and substituting into
(\ref{eq65},\ref{eq66}), getting:

\begin{eqnarray}
\left( E_{m}-E_{0}\right) a_{m}+{\sum_{m^{\prime }}}a_{m^{\prime }}\left[
W_{m,m^{\prime }}+{\sum_{\beta }}\frac{W_{m,\beta }W_{\beta ,m^{\prime }}}{
E_{0}-E_{\beta }}\right]  \nonumber \\
+{\sum_{\alpha }}b_{\alpha }\left[ W_{m,\alpha }+{\sum_{\beta }}\frac{
W_{m,\beta }W_{\beta ,\alpha }}{E_{0}-E_{\beta }}\right] =0  \label{17}
\end{eqnarray}

\begin{eqnarray}
\left( E_{\alpha }-E_{0}\right) b_{\alpha }+{\sum_{m^{\prime }}}a_{m^{\prime
}}W_{\alpha,m^{\prime }}  \nonumber \\
+{\sum_{\alpha ^{\prime }}}b_{\alpha ^{\prime }}\left[ W_{\alpha ,\alpha
^{\prime }}+{\sum_{\beta }}\frac{W_{\beta ,\alpha ^{\prime }}W_{\alpha
,\beta }}{E_{0}-E_{\beta }}\right] =0  \label{18}
\end{eqnarray}
Thus we see that the r\^{o}le of 6-body states is just to renormalize the
interaction between 2-body and 4-body ones, and for the rest they may be
forgotten about. If $E_{0}$ is outside the continuum of excitations, as we
shall show below, the corrections are finite, and experience with clusters
suggests that they are small. Had we included 8-body excitations, we could
have eliminated them by solving the system for their coefficients and
substituting, thus reducing to the above problem with further
renormalizations. In principle, the method applies to all the higher order
interactions, and we can recast our problem as if only 2 and 4-body states
existed. Again, the $W_{\alpha^{\prime } ,\alpha }$ term can be eliminated
by taking linear combinations of the $\alpha $~ states: when this is done,
we get a self-energy correction to $E_{\alpha }$ and a mixing of
the $W_{m,\alpha }$ vertices. The equations become

\begin{equation}
\left( E_{m}-E_{0}\right) a_{m}+{\sum_{m^{\prime }}}a_{m^{\prime
}}W_{m,m^{\prime }}+{\sum_{\alpha }}b_{\alpha }W_{m,\alpha }=0  \label{19}
\end{equation}

\begin{equation}
\left( E_{\alpha }-E_{0}\right) b_{\alpha }+{\sum_{m^{\prime }}}a_{m^{\prime
}}W_{\alpha ,m^{\prime }}=0  \label{20}
\end{equation}
Solving for $b_{\alpha }$ and substituting in the first equation we exactly
decouple the 4-body states as well. The eigenvalue problem is now 
\begin{equation}
\left( E_{0}-E_{m}\right) a_{m}=\sum_{m^{\prime}} a_{m^{\prime }}\left\{
W_{m,m^{\prime }}+\left\langle m|S[E_{0}]|m^{\prime }\right\rangle \right\} , 
\label{schro}
\end{equation}
where

\begin{equation}
\left\langle m|S\left[ E_{0}\right] |m^{\prime }\right\rangle ={\sum_{\alpha
}}\frac{<m|W|\alpha ><\alpha |W|m^{\prime }>}{E_{0}-E_{\alpha }}.  \label{eiv}
\end{equation}

Equation (\ref{schro}) is of the form of a Schr\"{o}dinger equation with eigenvalue
$E_{0}$ for pairs with an effective interaction $W+S$. Then we interpret
$a_{m}$ as the wave function of the dressed pair, which is acted upon by an
effective Hamiltonian $\tilde{H}$. The change from the full many-body H to $
\tilde{H}$ is the canonical transformation which generalizes the one of the
previous Section to all orders. In the new picture, the holes interact through an
effective vertex with infinitely many contributions, some of which are shown in
Figure 3. The linked contributions represent repeated exchange of electron-hole
pairs, and may contain self-energy insertions; all these contributions make up
the effective interaction. The unlinked diagrams are pure self-energy. Thus, the
scattering operator $S$ is of the form
$S=W_{eff}+F,$ where
$W_{eff}$ is the effective interaction between dressed holes, while $F$ is a
forward scattering operator, diagonal in the pair indices $m$ ,$m^{\prime }$ which
accounts for the self-energy corrections of the one-body propagators: it is
evident from (\ref{schro}) that it just redefines the dispersion law $E_{m}$, and,
essentially, renormalizes the chemical potential. Therefore $F$ must be
dropped, as in Cooper theory\cite{kn:kittel} and in the last
Section.
\begin{figure}
\begin{center}
	\epsfig{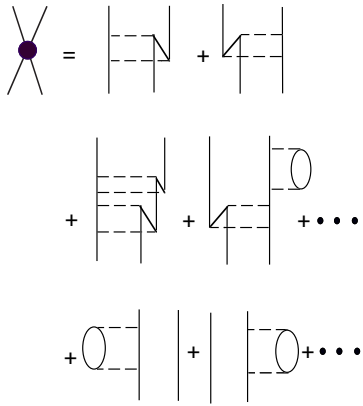}\caption{\footnotesize{
Some terms of the expansion of the effective hole-hole vertex (blob) defined by 
$\tilde{H}$. The dashed lines represent $W$ interactions. Linked diagrams belong
to $\tilde{W}_{eff}$, and unlinked ones to $F$.}}
\end{center} 
\end{figure}
So, the effective Schr\"{o}dinger equation for the pair reads

\begin{equation}
\left( H_{0}+W+W_{eff}\right) |a>=E_{0}|a>  \label{es}
\end{equation}
and we are interested in the possibility that $E_{0}=2E_{F}-| \Delta|
$, with a positive binding energy $| \Delta | $ of the
pair. Any other pairing mechanism not considered here, like off-site
interactions, inter-planar coupling and phonons, can be included as 
an extra contribution to $W_{m^{\prime },m}$ which  just adds to $W_{eff}$.

We emphasized the fact that the canonical transformation is exact because in
this way our argument does not require U/t to be small. In the 
numerical calculations below, we shall drop $W$ and approximate $W_{eff}$  
neglecting  the 6-body and higher excitations; in other terms, we shall compute the bare
quantities. At least the structure of the solution is exact when
expressed in terms of renormalized matrix elements, and the process 
can be systematically improved. The $\alpha $ states are
those of Equation (\ref{alfas}) and the interaction matrix element is
given in Equation (\ref{me}). Working out (\ref{eiv}) we find that the product in the numerator  yields 4
terms; two are proportional to $\delta (p-s)$ and belong to $F$, while the cross
terms yield identical contributions to $W_{eff}$. Using 
(\ref{os1},\ref{os2}), we obtain the
effective interaction between W=0 pairs:

\begin{eqnarray}
\left\langle \Phi _{\eta }\left[ p\right] \right| W_{eff}\left| \Phi _{\eta
}\left[ s\right] \right\rangle =\nonumber\\
4{\sum_{R\in C_{4v}}}\chi
^{\left( \eta \right) }\left( R\right) {\sum_{k}^{occ}}
\theta \left( \varepsilon \left( Rs+p+k\right) -E_{F}\right)
\nonumber\\
  \frac{U\left( Rs+p+k,-p,Rs,k\right)U\left( p,k,Rs+p+k,-Rs\right)}
{\varepsilon \left( Rs+p+k\right) -\epsilon \left( k\right)
+\epsilon \left( s\right) +\epsilon \left( p\right)-E_{0} }
 \label{weffective}
\end{eqnarray}
The sum is over occupied $k$ with empty $Rs+p+k$. Note that $W_{eff}$
does not depend on the sign of $U$. We may see that the perturbation 
theory of Section V yields the arithmetical mean of the two 
unperturbed limits  
$E_{0}\rightarrow 2\varepsilon\left( p\right) $ and $E_{0}\rightarrow 2\varepsilon
\left( s\right) $. The diagonal elements of (\ref{weffective}) are 
also clearly related to the $\Delta^{\left(2 \right)} $ expression derived from
perturbation theory for the fully symmetric clusters\cite{kn:cb5}, 
which are special cases of the present theory. The expressions 
(\ref{clust}),(\ref{wfi}) and (\ref{weffective}) are characterized 
by the symmetry-induced quantum mechanical interference of several terms. The sum can be positive 
or negative depending on the Irrep. 
This interference produces a partial cancellation, and the absolute value of the 
result is typically much smaller than individual contributions. This 
explains why the interaction is dynamically small for W=0 pairs:  
they  have no direct interaction, and because of the interference the effective interaction is 
reduced compared to what one could expect by a rough 
order-of-magnitude estimate. However, the presence of the $\theta$ 
functions and the anisotropy of the integrands prevent a total 
cancellation. 

In (\ref{clust}), the vanishing 
of the denominators is prevented by the discrete spectrum of the 
cluster; however, when applying (\ref{wfi}) to the plane the integrand 
is singular. This 
complication disappears in the full expression (\ref{weffective}) 
since we are  interested in bound pairs; then, $E_{0}$ is below the 
continuum and the denominators never vanish. 

 The $s$ and $p$ indices run over $1/8$ of the BZ. We denote
such a set of empty states $e/8$, and cast the result in the form of a
(Cooper-like) Schr\"{o}dinger equation

\begin{equation}
2\varepsilon \left( k\right) a\left( k\right) +\stackrel{e/8}{
\sum_{k^{\prime }}}W_{eff}\left( k,k^{\prime }\right) a\left( k^{\prime
}\right) =E_{0}a\left( k\right)  \label{equazione}
\end{equation}
for a self-consistent calculation of $E_{0}$ (since $W_{eff}$ depends on the
solution). Let $N_{C}$ be the number of cells in the crystal. The $U$ matrix
elements (\ref{ug},\ref{uc}) scale as $N_{C}^{-1}$ and therefore $W_{eff}$ scales in the same
way. For an infinite system, $N_{C}\rightarrow \infty $ , this is a well
defined integral equation. The existence of a clear-cut continuum 
limit means that the problem is well posed, but this equation is very difficult to treat
analytically. The interaction matrix is complicated by the Umklapp
discontinuities, other discontinuities come from the limitations to occupied or
empty states, and there are several independent variables since everything is
anisotropic. Therefore, we must resort to a numerical treatment. For the sake of
simplicity, we shall neglect the minor contributions from the higher bands and
consider the dominant intra-band processes, in which empty states belong to the
bonding band.

\section{Supercell Calculations}

Several good supercell calculations devoted to the problem of pairing  have been reported to 
date\cite{kn:dag}, but no conclusive evidence was reached, because of 
the difficulty of dealing with size effects. Any brute force approach 
will miss the result by far, because size effects are quite 
important, or if you wish, the minimum size useful for simulating the plane is enormous.
Now we have the advantage of starting from an analytic result 
(Equation (\ref{weffective})) which holds for the full plane. However 
the analytic solution is difficult, so we must discretise the 
problem. The best way is to work with a supercell of $N_{SC}\times
N_{SC}=N_{C}$ cells, with periodic boundary conditions. The advantage 
remains, however, because Equation (\ref{weffective}), however hard, is easy compared 
to direct diagonalisation.  We are able to
solve the problem in a virtually exact way for $N_{C}=30$ or $40$, while the
calculation becomes somewhat hard with larger supercells.

We loooked for triplet bound states, but we did not find them, because, as in the
clusters, $W_{eff}$ is repulsive for triplets. Therefore the rest of the analysis
is concerned with the W=0 singlets, which do show pairing, as we demonstrate below. 

The $\Delta <0$ behaviour
that we obtain in supercell calculations is in line with our previous findings in
small clusters(\cite{kn:cb1},\cite{kn:cb5},
\cite{kn:cb2},\cite{kn:cb3},\cite{kn:cb4}). The supercells have
periodic boundary conditions instead of the open boundary conditions we were using
previously and are very much larger systems than those that one can afford by
exact diagonalisation.  However, we are interested in the asymptotic behavior for
$N_{C}\rightarrow\infty $ and  $\Delta$ depends on $U^{\prime }s$ and $N_{C}$ in
a peculiar way, within the range of attainable supercell sizes. It increases with 
$N_{C}$ for large $U^{\prime }s$ and drops for small $U^{\prime }s$;  we
must understand the dependence in order to be able to make reliable
extrapolations.
  Cooper\cite{kn:kittel} assumed a constant
negative interaction
$-V_{C}$ within a distance $\omega_{D}$ from $E_{F}$, where $\omega_{D}$ is the
Debye frequency. His result is well known:
\begin{equation}
|\Delta_{C}|=\frac{2 \omega_{D}}{e^{\frac{1}{V_{C}\rho_{F}}}-1}
\label{coop}
\end{equation}
where $\rho_{F}$ is the density of states at the Fermi level. The orders of
magnitude are $\omega_{D}\approx 25$  $meV$,$\rho_{F}\approx 0.3$$ eV^{-1}/cell$
and $V_{C}\approx 1eV/cell$, and $|\Delta_{C}|$ is in the meV range.
 Here, no $\omega_{D}$ is involved, and all the empty states contribute;
however, we recall that all the distinct $W=0$ pairs may be labeled from 
the $k$ points in the $e/8$ set.  We define
the Uniform Interaction Model (UIM) in which a constant negative interaction $-V$
prevails for $k$ and $ k^{\prime }$ in $e/8$. Both the Cooper and the UIM 
models can be solved in supercell calculations, and show the same qualitative
behavior with increasing $N_{C}$. For small supercells, the binding energy is of the same
order as V, while for large cells it tends to the asymptotic value
$-\Delta_{asympt}$, which may be very different. This is similar to the behavior we
observe in the Hubbard model. Moreover, we can compute 
$\Delta $ for a given filling and $N_{SC}$ according to (\ref{equazione}), and
 determine the value of $V$ which gives the same $\Delta $ in the UIM.
 This will be $ V_{eff} $, that is, the effective $V$ of our theory,
characterizing the strength of the attraction by a single quantity. Fortunately,
$V_{eff}$ does not depend too much on the supercell size,and is seen to converge
with increasing $N_{SC}$. 

We use as input data the current estimates (in eV) t=1.3, $\varepsilon _{p}$
=3.5, $\varepsilon _{d}$=0, $U_{p}=6s$, $U_{d}=5.3s$, where s is a scale
factor induced by renormalization. At s=2.121, with $E_{F}$=-1.3 eV, we get 
for $^{1}B_{2}$ pairs the results shown in Table II.
Here, $n_{tot}$ is the filling and $V_{eff}$ is derived
by comparison with UIM calculations. We see that although
$\Delta|$ decreases monotonically with increasing supercell size up
to $N_{SC}=40$, $V_{eff}$ is fairly stable and is clearly {\em not} dropping to zero;
it corresponds to $|\Delta _{asympt}|$ values of about 20 meV.

 In addition,the UIM can be solved in the thermodynamic limit; writing
$E_{0}=2E_{F}-|\Delta| $ we can estimate $\Delta _{asympt}$ for $ N_{C}\rightarrow
\infty $ . Indeed, the Cooper-like Schr\"{o}dinger equation  
(\ref{equazione}) with
$W_{eff}=-\frac{V}{N_{C}}$, $V>0$, leads to

\begin{equation}
\frac{8}{V}=\int_{E_{F}}^{0}\frac{d\epsilon \rho \left( \varepsilon \right) 
}{2\left( \varepsilon -E_{F}\right) +|\Delta _{asympt}|}  \label{32}
\end{equation}

where $\rho $ is the density of states. This is readily solved numerically. The
dependence of $\Delta$ on V is qualitatively similar to the dependence of
$\Delta_{C}$ on $V_{C}$, and good fits can be obtained using the Cooper formula
with effective $\omega_{D}$ and $\rho_{F}$ parameters, like those shown in Table
III.

\section{Doping dependence}
In Table IV, we consider $^{1}B_{2}$ pairs at $E_{F}=-1.2$ eV (with the current
parameters, half filling corresponds to $E_{F}=-1.384$ eV, so the filling we are
considering is $n_{tot}\approx1.3$ at $N_{SC}=30$). We report values of $\Delta$
and $V_{eff}$ as obtained from supercell calculations at $N_{SC}$=12,20 and 30
for 3 values of the scale factor s. The extrapolation to the infinite supercell
is obtained from the UIM with the $V_{eff}$ value obtained at $N_{SC}=30$.
We see that the relatively mild $N_{SC}$ dependence of $V_{eff}$ supports the use
of the UIM to extrapolate the results to the thermodynamic limit, and there is a
clear indication of pairing with sizable binding energies. The $s$
  dependence of 
$V_{eff}$ is roughly linear, while $\Delta$ depends exponentially on $s$.
Table V presents the results for  $^{1}B_{2}$ pairs at $E_{F}=-1.1$ eV (
$n_{tot}\approx1.4$ at $N_{SC}=30$). The trend is similar, but $V_{eff}$ is seen
to increase with doping.
Tables VI and VII show the results for the $^{1}A_{2}$ pairs. These are seen to
lead to bound states as well, with comparable $\Delta$ values; the trend with
doping is opposite, however, and the gap is nearly closing at 
$E_{F}=-1.1 eV$.
A necessary condition for superconducting flux quantization is that two kinds of
pairs of similar binding energy and different symmetries exist\cite{kn:cbs}. A similar conclusion was reached independently by 
other Authors\cite{kn:schu}. Moreover, evidence of mixed $(s+id)$ 
symmetry for the pairing state has been amply reported in 
angle-resolved photoemission studies \cite{shen}. This remark leads to the prediction that in
this model superconductive pairing disappears with increasing $n_{tot}$, while
a different sort of pairing prevails; in reality the Cu-O plane
prefers to distort at excessive doping, and in a distorted plane the present mechanism,
based on symmetry, could be destroyed.

\section{Conclusions}
We  propose a general framework for the effective interaction between two
holes, based on the three-band Hubbard model but ready to include extra
interactions as well.  An effective Hamiltonian can in principle be obtained by a
systematic canonical transformation including any kind of virtual intermediate
states. We obtained the closed-form analytic expression of the effective
interaction including 4-body virtual states. This describes repeated 
exchange of an electron-hole pair. The argument does not depend on perturbation 
theory, and the equations retain their form, with renormalized 
parameters, at all orders. The previous exact-diagonalisation results 
of cluster calculations are special cases. The resulting integral Equation 
(\ref{equazione}), with the effective interaction(\ref{weffective}), is
{\it valid for the full plane}. Since an analytic treatment is 
prohibitive, we resort to a numerical treatment  by supercell calculations. 
Even so, the solution is hard.
Depending on the
parameters, extremely large supercells may be needed to obtain the convergence 
of the pair energy
$\Delta$ to the bulk limit; however, we find that precisely the same effect occurs in the
Uniform Interaction Model where a constant effective interaction $V$ is assumed. We
define  $V_{eff}$ as the value of $V$ that inserted in the UIM yields the same
$\Delta$ in supercell calculations as our integral equation. Since  $V_{eff}$
 converges to the bulk value much more readily, we are able to go to the
asymptotic limit, and to show the instability of the Fermi liquid in the model
at hand.  $\Delta _{asympt}$ values in the range of several tens to a few
hundreds of meV are obtained if we multiply the $U$ parameters by a scale factor s
which is somewhat larger than 1. The values $U_{d}=5.3 $ eV, $U_{p}=6. $ eV  differ appreciably 
from other literature estimates\cite{mcm}, and must depend on the 
compound and doping. For example, in La$_{2}$CuO$_{4}$, $U_{p}$=4 eV and 
$U_{d}$=10.5 eV have been recommended \cite{hyb}. However, since the screening excitations are
explicitly accounted for in the Hamiltonian, it is reasonable that the input
$U^{\prime }$s must be somewhat larger than the fully screened interaction.
Moreover, contributions from phonons and other mechanisms can be included 
as additive terms of $W$, and must be relevant for a comparison with experiment.
We find that
$^{1}A_{2}$ pairs are more tightly bound close to half filling, but $^{1}B_{2}$
pairs are favored when the filling increases. We remind the reader 
here that these symmetry labels are not absolute, but depend on the 
choice of a gauge convention\cite{nota}.
We get attraction and pairing at
all fillings, but the binding energy of the $^{1}A_{2}$ pairs drops by 
orders of magnitude as the filling increases; thus, there is no chance 
of superconducting flux quantization too far from half filling. So, 
we do not predict that superconductivity occurs outside some range of 
hole concentration. However, pairing is still there, even for large doping: 
since the present mechanism is 
driven by symmetry it works unless the system distorts. In fact, at 
excessive dopings the real superconductors develop stripes, and 
become normal metals.
The three-band Hubbard model might be too idealized to allow a detailed
comparison with experiments; however we stress that  the approach presented is far
more general than the model we are using, and can be applied to more realistic
Hamiltonians. This is the main result of the present paper.

This work has been supported by the Is\-ti\-tu\-to Na\-zio\-na\-le di Fisica
della Materia. We gratefully acknowledge A. Sa\-gnot\-ti, Universit\`{a} di
Roma Tor Ver\-ga\-ta, for useful and stimulating discussions.


\newpage

\begin{table}[tbp]
\begin{center}
\begin{tabular}{lclclclclclcl}
C$_{4v}$ & E & C$_{2}$ & 2C$_{4}$ & 2$\sigma $ & 2$\sigma ^{\prime }$ &  & 
&  &  &  &  &  \\ 
A$_{1}$ & 1 & 1 & 1 & 1 & 1 &  &  &  &  &  &  &  \\ 
A$_{2}$ & 1 & 1 & 1 & -1 & -1 & $R_{z}$ &  &  &  &  &  &  \\ 
B$_{1}$ & 1 & 1 & -1 & 1 & -1 & $x^{2}-y^{2}$ &  &  &  &  &  &  \\ 
B$_{2}$ & 1 & 1 & -1 & -1 & 1 & $xy$ &  &  &  &  &  &  \\ 
E & 2 & -2 & 0 & 0 & 0 & $\left( x,y\right) $ &  &  &  &  &  & 
\end{tabular}
\end{center}
\caption{The Character Table of the $C_{4v}$ Group}
\label{Table I}
\end{table}

\narrowtext
\begin{table}[tbp]
\begin{center}
\begin{tabular}{lclclclcl}
$N_{SC}$ & $n_{tot}$ & $-\Delta \left( meV\right) $ & $V_{eff}\left(
eV\right) $ & $\Delta _{asympt}\left( meV\right) $ &  &  &  &  \\ 
18 & 1.13 & 121.9 & 7.8 & 41.6 &  &  &  &  \\ 
20 & 1.16 & 42.2 & 5. & 9.0 &  &  &  &  \\ 
24 & 1.14 & 59.7 & 7. & 28.9 &  &  &  &  \\ 
30 & 1.14 & 56. & 5.7 & 13.2 &  &  &  &  \\ 
40 & 1.16 & 30.5 & 6.6 & 23.4 &  &  &  & 
\end{tabular}
\end{center}
\caption{Binding Energy of $^{1}B_{2}$ Pairs in supercells.}
\label{supercells}
\end{table}

\narrowtext
\begin{table}[tbp]
\begin{center}
\begin{tabular}{lclclcl}
$E_{F}(eV)$ & $\omega_{D}(eV)$ & $\rho_{F}(eV^{-1}/Cell) $ \\ 
-1.35 & 0.4545 & 0.0486   \\ 
-1.3 & 0.492 & 0.0415   \\ 
-1.2 & 0.520 &0.0338   \\ 
-1.1 & 0.5237 & 0.02905   \\ 

\end{tabular}
\end{center}
\caption{Effective Cooper parameters for fitting the V dependence of
$\Delta$ at several $E_{F}$ values.}
\label{Effective Cooper}
\end{table}

\narrowtext
\begin{table}[tbp]
\begin{center}
\begin{tabular}{lclclclclc|c|c|}

&$s=$&$\sqrt{2}$&$s=$&$\frac{3}{\sqrt{2}}$&$s=$&$2\sqrt{2}$\\
$N_{SC}$ & $-\Delta\left(meV\right)$ & $V_{eff} $& $-\Delta\left(meV\right)$ &
$V_{eff} $& $-\Delta\left(meV\right)$ & $V_{eff} $ \\
  12 &126 &6.75& 251.&10.&469.&14.23  \\
  20 & 63. &6.9 &135.&10.6&320.5&15.9  \\ 
30 & 27. &6.3& 198. &13.8&460.&20.8   \\ 
$\infty$ & 8.8 &6.3& 131.&13.8& 333.&20.8  
\end{tabular}
\end{center}
\caption{Data for $^{1}B_{2}$ pairs at $E_{F}=-1.2\,eV$; $V_{eff}$ is in 
eV.}
\label{Numerical Results1.}
\end{table}

\narrowtext
\begin{table}[tbp]
\begin{center}
\begin{tabular}{lclclclclc|c|c|}

&$s=$&$\sqrt{2}$&$s=$&$\frac{3}{\sqrt{2}}$&$s=$&$2\sqrt{2}$\\
$N_{SC}$ & $-\Delta\left(meV\right)$ & $V_{eff} $& $-\Delta\left(meV\right)$ &
$V_{eff} $& $-\Delta\left(meV\right)$ & $V_{eff} $ \\
  12 &163. &9.& 409.&15.7&688.&22.1  \\
  20 & 151. &10.8 &371.&17.6&635.5&24.3  \\ 
30 & 80. &10.3& 270. &17.66&520.&24.9  \\ 
$\infty$ & 35.6 &10.3& 170.8&17.66& 362.&24.9  
\end{tabular}
\end{center}
\caption{Data for $^{1}B_{2}$ pairs at $E_{F}=-1.1\,eV$; $V_{eff}$ is in 
eV.}
\label{1b2}
\end{table}

\narrowtext
\begin{table}[tbp]
\begin{center}
\begin{tabular}{lclclclclc|c|c|}

&$s=$&$\sqrt{2}$&$s=$&$\frac{3}{\sqrt{2}}$&$s=$&$2\sqrt{2}$\\
$N_{SC}$ & $\Delta\left(meV\right)$ & $V_{eff} $& $\Delta\left(meV\right)$ &
$V_{eff} $& $\Delta\left(meV\right)$ & $V_{eff} $ \\
  12 &141. &7.2& 274.&10.43&423.&13.4  \\
  20 & 46.2 &6.5 &96.&9.1&158.2&11.4  \\ 
30 & 25.9 &6.23& 64.8 &8.8&131.&11.6   \\ 
$\infty$ & 8.35 &6.23& 34.&8.8& 81.7&11.6  
\end{tabular}
\end{center}
\caption{Data for $^{1}A_{2}$ pairs at $E_{F}=-1.2\,eV$; $V_{eff}$ is in 
eV.}
\label{Numerical Results2.}
\end{table}

\narrowtext
\begin{table}[tbp]
\begin{center}
\begin{tabular}{lclclclclc|c|c|}

&$s=$&$\sqrt{2}$&$s=$&$\frac{3}{\sqrt{2}}$&$s=$&$2\sqrt{2}$\\
$N_{SC}$ & $\Delta\left(meV\right)$ & $V_{eff} $& $\Delta\left(meV\right)$ &
$V_{eff} $& $\Delta\left(meV\right)$ & $V_{eff} $ \\
  12 &72. &5.5& 144.&8.4&228.&11.  \\
  20 & 10. &2.4 &26.5&4.38&61.&6.78  \\ 
30 & 5.48 &2.5& 19. &5.8&45.&8.16  \\ 
$\infty$ & 1.5 &2.5& 2.6&5.8& 14.4&8.16  
\end{tabular}
\end{center}
\caption{Data for $^{1}A_{2}$ pairs at $E_{F}=-1.1\,eV$; $V_{eff}$ is in 
eV.}
\label{Numerical Results3.}
\end{table}

\bigskip


\begin{references}

\bibitem{kn:bm}  J.G. Bednorz and K.A. Muller, Z. Phys. B{\bf 64} (1986) 189.

\bibitem{kn:pietro}  L. Pietronero and S. Str\"{a}sser, Europhys. Lett{\bf 18%
}, 627 (1992).

\bibitem{kn:iad}  G. Iadonisi,M. Chiofalo, V. Cataudella and D. Ninno, Phys.
Rev. B{\bf 48}, (1993) 12966 and references therein.

\bibitem{kn:bob}  J.R.Schrieffer, X.C.Wen and S.C.Zhang, Phys. Rev. Letters,%
{\bf \ 60}, 944 (1988).

\bibitem{kn:pw}  P.W. Anderson, Science,235,(1987),1196; see also {\em The
Hubbard Model}, Ed. D. Baeriswyl et al., Plenum Press, New York, 1995, page
217.

\bibitem{kn:sus}  O.P. Sushkov, Phys. Rev. B{\bf 54}, 9988 (1996) and
reference therein; V.V. Flambaum, M.Yu. Kuchiev and O.P. Sushkov, 
Physica C {\bf 235-240}, 2218 (1994).

\bibitem{kn:hirsch}  J.E.Hirsch,S.Tang, E.Loh Jr, and D.J.Scalapino, Phys.
Rev.Letters{\bf \ 60}, 1668 (1988); Phys. Rev. B{\bf 39}, 243 (1989).

\bibitem{kn:bal}  C.A.Balseiro, A.G.Rojo, E.R.Gagliano, and B.Alascio,
Phys.Rev.B {\bf 38}, 9315 (1988).

\bibitem{kn:jones} D.H. Jones and S.J.Monaghan, J. Phys. Condens. 
Matter, {\bf 1} (1989) 1843.

\bibitem{kn:tink}S. Mazumdar, F. Guo, D. Guo, K.C. Ung and J. Tinka Gammel,
Proc. of the Discussion Meeting on Strongly Correlated Electron Systems in
Chemistry, Bangalore, India (Springer Verlag, Berlin 1996).

\bibitem{kn:cb1}  M. Cini and A. Balzarotti, Il Nuovo Cimento D {\bf 18}, 89
(1996).

\bibitem{kn:saw}M.A. Van Veenendal and G.A. Sawatzky, Phys. Rev. 
Lett. {\bf70}, 2459 (1993).

\bibitem{kn:sch}M. Schluter, in {\em Superconductivity and 
Applications}, edited by H.S. Kwok {\em et al.} (Plenum, New York, 
1990),p.1.

\bibitem{kn:schu} H.-B. Sch\"{u}ttler, C. Gr\"ober,H.G. Evertz and W. 
Hanke, cond-mat/9805133

\bibitem{kn:cb5}  M. Cini and A. Balzarotti, Phys. Rev. B{\bf 56}, 1, 14711
(1997).

\bibitem{kn:cb2}  M. Cini and A. Balzarotti, J. Phys. Condens. Matter{\bf 8}
, L265 (1996).

\bibitem{kn:cb3}  M. Cini and A. Balzarotti, Solid State Commun.{\bf 101},
671 (1997).

\bibitem{kn:cb4}  M. Cini, A. Balzarotti, J. Tinka Gammel and A. R. Bishop,
Nuovo Cimento {\bf 19 D}, (1997) 1329.

\bibitem{kn:cbs}  M. Cini, A. Balzarotti and G. Stefanucci, submitted
for publication.
\bibitem{kn:csbcond}  M. Cini, A. Balzarotti and G. Stefanucci,  
cond-mat/9808209 and submitted for publication.
\bibitem{kn:san} Raimundo R. dos Santos, cond-mat/9502043.

\bibitem{kn:kittel} L.Cooper, Phys. Rev. B{\bf 104}, 1189
  (1956); C. Kittel, {\em
Quantum Theory of Solids}, John Wiley and Sons, New York (1963) Chapter 8.

\bibitem{kn:dag} E. Dagotto, Rev. Mod. Phys. {\bf 66 1324}, (1994) 763.

\bibitem{shen} Z.X. Shen, W.E. Spicer, D.M. King, D.S. Dessau and
B.O. Wells, Science {\bf 267}, 343 (1995) and References therein. 

\bibitem{mcm} A.K. McMahan, R.M. Martin and  S. Satpathy, Phys. Rev. 
B {\bf 38}, 6650 (1988).

\bibitem{hyb} M. S. Hybertsen, E.B. Stechel, M. Schluter and D. 
R.Jennison, Phys. Rev.  B{\bf 41}, 11068 (1990). 

\bibitem{nota} In comparing these results with those 
 of various calculations from other 
Authors, one should bear in mind that in the three-band Hubbard model for the 
Cu-O plane different conventions have been widely used in the 
literature for the phase 
of orbitals, which are equally possible because of the gauge 
invariance of the theory. We are using  a model Hamiltonian 
(\ref{h0}), which is simplest because a Cu ion is bonded to the 
nearest neighbor O by identical hopping integrals $t$; however, various 
Authors prefer an alternating bond convention, with hopping integrals 
changing sign for a $90^{0}$ rotation. With to the choice of the 
gauge, the symmetry labels vary, according to the multiplication 
table of Irreps of the $C_{4v}$ Group.

\end{references}
\end{document}